\documentclass{article}

\usepackage{arxiv}

\usepackage[utf8]{inputenc} % allow utf-8 input
\usepackage[T1]{fontenc}    % use 8-bit T1 fonts
\usepackage{hyperref}       % hyperlinks
\usepackage{url}            % simple URL typesetting
\usepackage{booktabs}       % professional-quality tables
\usepackage{amsfonts}       % blackboard math symbols
\usepackage{nicefrac}       % compact symbols for 1/2, etc.
\usepackage{microtype}      % microtypography
\usepackage{lipsum}		% Can be removed after putting your text content
\usepackage{graphicx}
\usepackage{doi}
\usepackage{multirow}
\usepackage{caption}
\usepackage{subcaption}
\usepackage{booktabs}

\usepackage[sort, numbers, square]{natbib}

\usepackage[acronym,nohypertypes={acronym, notation}]{glossaries}
\makeglossaries
\newacronym{MCTS}{MCTS}{Monte-Carlo Tree Search}
\newacronym{DFPN}{DFPN}{Depth-First Proof-Number}
\newacronym{AZF}{AZF}{AiZynthFinder}

\newacronym[shortplural={RNNs}, longplural={Recurrent Neural Networks}]{RNN}{RNN}{Recurrent Neural Network}

\title{Mind the Retrosynthesis Gap:\\Bridging the divide between Single-step and Multi-step Retrosynthesis Prediction}

\date{} 					% Or removing it

\author{%
  Alan Kai Hassen\thanks{The following authors contributed equally: Alan Kai Hassen, Paula Torren-Peraire} \\
  Leiden Institute of Advanced Computer Science \\
  Leiden University\\
  Leiden, The Netherlands \\
  \texttt{a.k.hassen@liacs.leidenuniv.nl} \\
  \And
  Paula Torren-Peraire$^*$ \\
  Institute of Structural Biology\\
  Molecular Targets and Therapeutics Center \\
  Helmholtz Zentrum München \\
  Neuherberg, Germany\\
  \texttt{paula.torren@helmholtz-munich.de} \\
  \And
  Samuel Genheden \\
  Molecular AI \\ 
  Discovery Sciences, R\&D \\
  AstraZeneca \\
  Gothenburg, Sweden \\
  \texttt{samuel.genheden@astrazeneca.com} \\
  \And
  Jonas Verhoeven \\
  In-Silico Discovery and External Innovation \\
  Janssen Research \& Development \\
  Janssen Pharmaceutica N.V \\ 
  Beerse, Belgium \\
  \texttt{jverhoe9@its.jnj.com} \\
  \And
  Mike Preuss \\
  Leiden Institute of Advanced Computer Science \\
  Leiden University\\
  Leiden, The Netherlands \\
  \texttt{m.preuss@liacs.leidenuniv.nl} \\
  \And
  Igor Tetko \\
  Institute of Structural Biology\\
  Molecular Targets and Therapeutics Center \\
  Helmholtz Zentrum München \\
  Neuherberg, Germany \\
  \texttt{igor.tetko@helmholtz-munich.de} \\
}

\renewcommand{\headeright}{}
\renewcommand{\undertitle}{}
\renewcommand{\shorttitle}{Mind the Retrosynthesis Gap}

\hypersetup{
pdftitle={Mind the Retrosynthesis Gap: Bridging the divide between Single-step and Multi-step Retrosynthesis Prediction},
pdfauthor={Alan Kai Hassen, Paula Torren-Peraire, Samuel Genheden, Jonas Verhoeven, Mike Preuss, Igor Tetko},
pdfkeywords={Computer-Aided Synthesis Planning, Retrosynthesis, Synthesis, Retrosynthesis Prediction, Benchmark, Monte Carlo Tree Search, Retro*, LocalRetro, MHNreact, Chemformer, Template-Based, AiZynthFinder},
}

\begin{document}
\maketitle

\begin{abstract}
Retrosynthesis is the task of breaking down a chemical compound recursively step-by-step into molecular precursors until a set of commercially available molecules is found. Consequently, the goal is to provide a valid synthesis route for a molecule. As more single-step models develop, we see increasing accuracy in the prediction of molecular disconnections, potentially improving the creation of synthetic paths. Multi-step approaches repeatedly apply the chemical information stored in single-step retrosynthesis models. However, this connection is not reflected in contemporary research, fixing either the single-step model or the multi-step algorithm in the process. In this work, we establish a bridge between both tasks by benchmarking the performance and transfer of different single-step retrosynthesis models to the multi-step domain by leveraging two common search algorithms, Monte Carlo Tree Search and Retro*. We show that models designed for single-step retrosynthesis, when extended to multi-step, can have a tremendous impact on the route finding capabilities of current multi-step methods, improving performance by up to +30\% compared to the most widely used model. Furthermore, we observe no clear link between contemporary single-step and multi-step evaluation metrics, showing that single-step models need to be developed and tested for the multi-step domain and not as an isolated task to find synthesis routes for molecules of interest.
\end{abstract}

% keywords can be removed
\keywords{Computer-Aided Synthesis Planning \and Retrosynthesis \and Synthesis \and Retrosynthesis Prediction \and Benchmark \and Monte Carlo Tree Search \and Retro* \and LocalRetro \and MHNreact \and Chemformer \and Template-Based \and AiZynthFinder}

\section{Introduction}
Artificial intelligence has greatly impacted the drug discovery pipeline by achieving human-like performance in the field of retrosynthesis \cite{seglerPlanningChemicalSyntheses2018}. Retrosynthesis is the task of breaking down a chemical compound recursively into molecular precursors until a set of commercially available building block molecules is found \cite{seglerPlanningChemicalSyntheses2018, coreyLogicChemicalSynthesis1989}. Consequently, the goal is to provide a valid synthesis route for a molecule. Potential applications of these synthesis routes are suggestions for medical chemists on how to produce a molecule of interest \cite{coleyMachineLearningComputerAided2018}, a foundation for autonomous chemistry \cite{coleyRoboticPlatformFlow2019}, and using synthesizability as part of De Novo Drug Design \cite{schneiderComputerbasedNovoDesign2005}.

The field of computational retrosynthesis prediction is separated into two different tasks \cite{schwallerMachineIntelligenceChemical2022}. In single-step retrosynthesis, the goal is to find the likely precursors or reactants for a given product. In multi-step retrosynthesis planning, the goal is to find viable synthesis paths over multiple reaction steps. 

The task of single-step retrosynthesis prediction is treated as a supervised learning task, commonly categorized as template-based or template-free \cite{dongDeepLearningRetrosynthesis2021}. Template-based approaches use manually curated or data-driven reaction templates \cite{thakkarArtificialIntelligenceAutomation2021}. These templates represent the general atoms and bond structures required to perform a reaction. Therefore the objective is to predict the most appropriate template to break down the molecule \cite{seidlImprovingFewZeroShot2022, chenDeepRetrosyntheticReaction2021a}. Template-free approaches are treated as a sequence prediction problem predicting one token of a chemical SMILES vocabulary at a time \cite{tetkoStateoftheartAugmentedNLP2020, irwinChemformerPretrainedTransformer2022}, drawing inspiration from natural language processing \cite{vaswaniAttentionAllYou2017}. Recently variations of these two approaches have emerged. In semi-template-based \cite{sachaMoleculeEditGraph2021a, wangRetroPrimeDiversePlausible2021a}, a molecule is first broken down into subparts, and then the subparts are rebuilt into chemically viable reactants. Lastly, though many models leverage the sequence-based SMILES notation, there are also attempts to utilize graph-based descriptors across these approaches, exploiting the advantages of a molecular graph \cite{chenDeepRetrosyntheticReaction2021a, tuPermutationInvariantGraphtoSequence2022}.

In comparison to single-step retrosynthesis, multi-step retrosynthesis planning focuses on researching novel route search algorithms using a fixed single-step model to identify retrosynthetic disconnections. The pioneering work in the field uses neural-guided \gls{MCTS} and a template-based approach to find synthesis routes \cite{seglerPlanningChemicalSyntheses2018}. Instead of assessing the state value in the search tree at run-time, alternative methods use oracle functions to guide the tree search. These methods include \gls{DFPN} search with edge cost, which combines classical \gls{DFPN} with a neural heuristic \cite{kishimotoDepthfirstProofnumberSearch2019}, and Retro*, which combines the A* path finding algorithm with a neural heuristic \cite{chenRetroLearningRetrosynthetic2020}. Newer approaches use a template-free model, either combining neural-guided \gls{MCTS} with reaction feasibility heuristics \cite{linAutomaticRetrosyntheticRoute2020} or directly using synthesizability heuristics combined with a forward synthesis model \cite{schwallerPredictingRetrosyntheticPathways2020}. Instead of using heuristics, self-play \cite{silverMasteringGameGo2017}, learning a value function by letting an algorithm play the game of synthesis against itself, is an additional investigated approach \cite{schreckLearningRetrosyntheticPlanning2019b, hongRetrosyntheticPlanningExperienceGuided2021, kimSelfImprovedRetrosyntheticPlanning2021}.

Multi-step approaches repeatedly apply the chemical information stored in single-step retrosynthesis models. However, the relationship between single-step models and multi-step approaches is not reflected in contemporary research, treating both tasks as distinct entities.  Even though multi-step algorithms require the use of single-step models, these single-step models are generally fixed. Similarly, single-step models are developed without evaluating their use in multi-step approaches. Therefore, an open question is how single-step retrosynthesis evaluation metrics translate to the multi-step domain \cite{schwallerMachineIntelligenceChemical2022} and, consequently, how single-step models affect the synthesis route finding capabilities as part of a multi-step algorithm. In this work, we establish a bridge between single and multi-step retrosynthesis tasks by benchmarking the performance and transfer of different single-step retrosynthesis models to the multi-step domain. We show the impressive impact of the single-step model on multi-step performance but, more importantly, a disconnection between contemporary single-step and multi-step evaluation metrics.

\section{Methods}
We select three state-of-the-art retrosynthesis single-step models to compare their performance in the multi-step domain. The model selection is based on dominant contemporary neural network approaches, i.e., contrastive learning, sequence-to-sequence, and graph-based encoding, considering their respective top-1 to top-50 performance on the USPTO-50k single-step retrosynthesis benchmark \cite{loweExtractionChemicalStructures2012, schneiderWhatWhatNearly2016}. Accordingly, the selected models are MHNreact \cite{seidlImprovingFewZeroShot2022}, a contrastive learning approach,  Chemformer \cite{irwinChemformerPretrainedTransformer2022}, a sequence-to-sequence approach, and LocalRetro \cite{chenDeepRetrosyntheticReaction2021a}, a graph-based approach. As an additional baseline, a template-based multi-layer perceptron approach \cite{thakkarDatasetsTheirInfluence2020, genhedenAiZynthFinderFastRobust2020}, drawing inspiration from \cite{seglerPlanningChemicalSyntheses2018}, is included since it is often used in multi-step retrosynthesis algorithms \cite{seglerPlanningChemicalSyntheses2018, kishimotoDepthfirstProofnumberSearch2019, chenRetroLearningRetrosynthetic2020, kimSelfImprovedRetrosyntheticPlanning2021}. Given that we aim to evaluate the capacity of these single-step retrosynthesis models in multi-step retrosynthesis planning, we use the model hyperparameters suggested in their respective publications (Appendix \ref{tab:ssm_hyperparam}) assuming the models are optimized for the single-step prediction task. The only exception is Chemformer, where we use beam size 50 to produce the single-step results and beam size 10, the publication default, for multi-step retrosynthesis planning.

To ensure the correct implementation of the single-step models and compare their single-step performance, we perform a 10-fold cross-validation by splitting the data into 80\% training, 10\% validation, and 10\% test splits for each fold. Each model is trained using the train split, training is monitored using the validation split, and the test split is used for final evaluation. All models use the same data split for each fold, and the data is preprocessed according to the specifications of each model.

Each single-step model is evaluated by measuring its accuracy and inference time. Single-step accuracy \cite{dongDeepLearningRetrosynthesis2021} is defined as the percentage of target compounds for which the model finds the ground-truth reactants within the top-k, $k \in \{1,3,5,10,50\}$, measuring the ability of the model to capture chemical reaction information. Inference time is defined as the time needed to produce retrosynthesis predictions for a set of molecules, measuring the ability of the model to provide predictions in a timely manner. In an ablation study, we measure the impact of the amount of evaluation data and batch size on the inference time. For the first, we measure the influence of doubling the evaluation data while using the default batch size (Appendix \ref{tab:batch_sizes}), analyzing the scalability of the model. For the second, we measure the impact of setting the batch size to 1, replicating the necessary conditions for a multi-step search algorithm that can only explore one molecule per instance (e.g. \cite{seglerPlanningChemicalSyntheses2018}).

The selected multi-step algorithms to evaluate the performance of the different single-step models are \gls{MCTS} \cite{seglerPlanningChemicalSyntheses2018}, which dynamically assesses the state values of the search tree at run-time, and Retro* \cite{chenRetroLearningRetrosynthetic2020}, which instead uses an A* path finding algorithm in combination with an oracle function. 
In the case of Retro*, we refrain from using the oracle function and rely only on the priors of the single-step model for initial cost estimation, given that the original oracle function is generally shown to have little impact \cite{trippReEvaluatingChemicalSynthesis2022} and is trained on USPTO data, which could cause information leakage. We defer from using a self-play algorithm since it would be necessary to retrain the self-play algorithm per problem instance, i.e., the set of used building blocks.
%In the case of Retro*, we refrain from using the oracle function and rely only on the priors of the single-step model for initial cost estimation, given that the original oracle function is trained on USPTO data, which could cause information leakage. We defer from using a self-play algorithm since it would be necessary to retrain the self-play algorithm per problem instance, i.e., the set of used building blocks.

In the first and second experiments, the search settings for \gls{MCTS} and Retro* are set to a time limit of 1800 seconds (30 minutes) and 200 algorithm iterations per molecule (Appendix \ref{tab:ms_hyperparam}), respectively. In a third experiment, the search settings for Retro* are set to a time limit of 28800 seconds (8 hours) to allow the single-step models to reach the maximum iteration limit (Retro*-extended) (Appendix \ref{tab:ms_hyperparam}), given their potential slow inference times. This third experiment is only conducted with Retro* because the algorithm does not need to do multiple single-step model calls to evaluate a tree-search state. Thus the algorithm is more likely to allow the single-step model to reach the iteration limit. In this case, a single-step model call refers to the suggestion of multiple candidate reactants given a product. In all cases, we search up to a maximum route length, or tree depth, of 7 and use the Zinc stock of 17,422,831 building blocks \cite{genhedenAiZynthFinderFastRobust2020}. All experiments are conducted by extending the open-source \gls{AZF} multi-step retrosynthesis framework \cite{genhedenAiZynthFinderFastRobust2020} to use alternative single-step models instead of the thus far implemented baseline template-based model.

To evaluate the performance of a single-step model within a multi-step setting, we measure the solvability of molecules when searching for a synthesis plan. Solvability is the percentage of test molecules for which a specific combination of search algorithm and single-step model can produce solved routes. A route is considered solved when all predicted leaf compounds are available within the building block stock. To further investigate the performance of single-step retrosynthesis models, we also analyze the average number of iterations carried out by the search algorithm, the average number of calls to the single-step model, and the average search time, calculated across the test compounds.

The data used for all experiments is USPTO-50k \cite{schneiderWhatWhatNearly2016, loweExtractionChemicalStructures2012}, a commonly used dataset within the single-step retrosynthesis field. The dataset consists of 50,016 unique products and their respective reactants, where the randomly split dataset contains 40,012 training reactions, 5,002 validation reactions, and 5,002 test reactions. The multi-step evaluation is conducted on the products of the test set.

Single-step retrosynthesis models are trained and benchmarked on one Tesla V100 32GB GPU. In comparison, multi-step retrosynthesis experiments are evaluated using a high-performance CPU cluster to facilitate the parallelization necessary to evaluate an extensive set of molecules in an appropriate time frame.
\section{Results}
\subsection{Single-step retrosynthesis prediction}
\begin{figure}[t]
    \centering
    \includegraphics[width=0.615\textwidth]{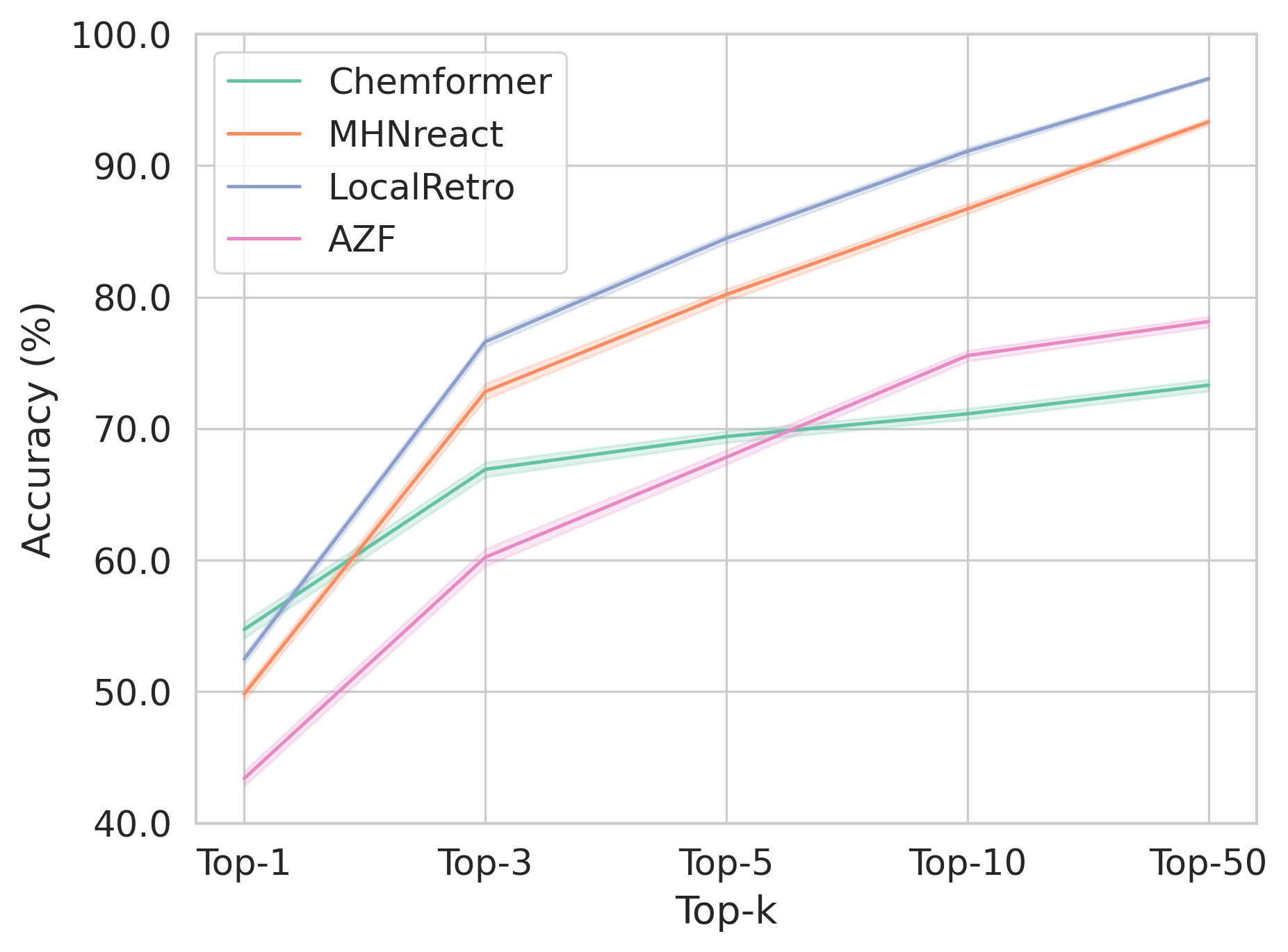} % reduced image size to 0.65 instead of 0.7
    \caption{Percentage of compounds for which single-step retrosynthesis models found the ground-truth reactants within the top-k (Accuracy) on USPTO-50k averaged across 10-fold cross-validation. The standard deviation over all folds is indicated by the colored error bands.}
    \label{fig:accuracy}
\end{figure}

We reproduce the performance of the selected single-step models using a 10-fold cross-validation with USPTO-50k. By averaging across the folds, we reproduce the results reported in Chemformer \cite{irwinChemformerPretrainedTransformer2022}, MHNreact \cite{seidlImprovingFewZeroShot2022}, and LocalRetro \cite{chenDeepRetrosyntheticReaction2021a} (Fig. \ref{fig:accuracy} and Appendix \ref{tab:accuracy}). For all models the data split has no discernable effect on the accuracy, shown by the small standard deviation across all folds. Additionally, we calculate the performance of the baseline model implemented in \gls{AZF} \cite{genhedenAiZynthFinderFastRobust2020} as a benchmark to compare the single-step models. 
For top-1 accuracy, Chemformer outperforms the other models, with an average accuracy of 54.7\% (± 1.1\%). LocalRetro and MHNreact follow this with 52.5\% (± 0.7\%) and 49.8\% (± 0.8\%) accuracy, respectively, and \gls{AZF} performs notably worse with an accuracy of 43.3\% (± 1.0\%). This pattern, however, is not maintained across the top-k measures. Accuracy noticeably ascends within the top-3 for all models, LocalRetro seeing a +24.1\% increase in accuracy to 76.6\% (± 0.6\%), similar to MHNreact with a +23.0\% increase to 72.8\% (± 1.0\%). \gls{AZF} has a +16.8\% increase in accuracy to 60.0\% (± 1.0\%), and Chemformer has the smallest gain in accuracy with an +11.2\% increase to 65.9\% (± 1.0\%). Within the top-50 predictions, LocalRetro shows 96.6\% (± 0.3\%) accuracy, followed by MHNreact with 93.3\% (± 0.4\%) accuracy, both showing similar profiles across the top-k. \gls{AZF} notably increases its performance across top-3 to top-10, giving 78.1\% (± 0.7\%) accuracy at top-50.  Surprisingly, Chemformer delivers the lowest accuracy in the top-50 of the models tested, with an accuracy of 73.3\% (± 0.3\%). Though Chemformer outperforms other models in top-1, it is less able to find the ground-truth reactants for the remaining products despite additional explored alternatives with higher top-k.

\begin{figure}[b]
    \centering
    \includegraphics[width=0.6\textwidth]{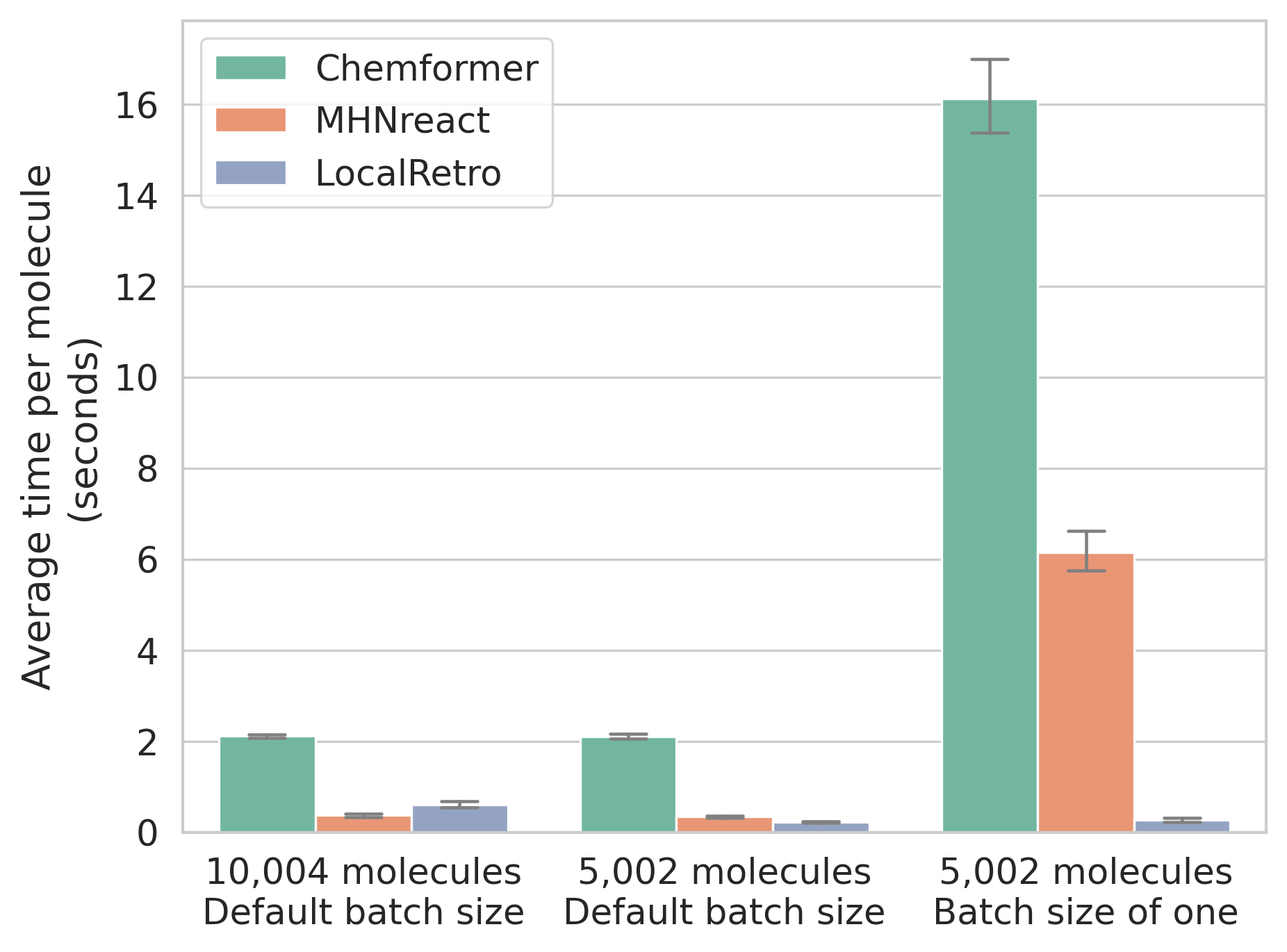} % reduced image size to 0.65 instead of 0.7
    \caption{Influence of data and batch size (Appendix \ref{tab:batch_sizes}) on inference time per molecule on USPTO-50k averaged across 10-fold cross-validation.}
    \label{fig:times}
\end{figure}
The influence of increased data and decreased batch size on the single-step model inference time is examined since single-step models typically evaluate in batches (Fig. \ref{fig:times}). Generally, there is a linear relationship when doubling the amount of inferred data. We observe that the inference time per molecule remains stable when the amount of test data increases, except for LocalRetro, which triples the inference time per molecule. 
In contrast, by decreasing batch size to one, we emulate the conditions of the model call within multi-step retrosynthesis planning. Chemformer and MHNreact both substantially increase their average inference time per molecule. For Chemformer, the increase in inference time is discernible, reaching eight times compared to the default batch size. MHNreact has the most marked increase in inference time, with the change from batch size 32 to 1 leading to 18x longer inference time per molecule. On the other hand, the inference time per molecule of LocalRetro is hardly affected by this change.

\subsection{Multi-step retrosynthesis planning}
Introducing the single-step models into the selected search algorithms, \gls{MCTS} generally performs worse than Retro* (Table \ref{tab:ms}). In detail, Retro* performs better in terms of solvability, number of explored routes, and solved routes per molecule across nearly all tested single-step models. The only exception is Chemformer, which produces more solved routes per molecule (\gls{MCTS}: 3.33, Retro*: 2.06) while using fewer model calls (\gls{MCTS}: 8.48, Retro*: 14.4). However, Chemformer with \gls{MCTS} still has a lower overall solvability (\gls{MCTS}: 44.3\%, Retro*: 53.4\%). In essence, it produces multiple solved routes for a smaller subset of solved molecules. Retro*-extended reaches or improves the result of Retro*, given that single-step models have more time for inference. In detail, Chemformer and MHNreact achieve higher performance using Retro*-extended, leveraging more single-step model calls. In comparison, the baseline \gls{AZF} model and LocalRetro do not utilize the added time with more single-step model calls as they already reach the 200 iteration limit within the 30 min time limit of Retro*, thus having similar performance using both settings.

For the overall best-performing search setting, Retro*-extended, the single-step model ranking in terms of solvability is LocalRetro (80.6\%), Chemformer (65.6\%), MHNreact (60.9\%), and \gls{AZF} (50.6\%). However, high solvability does not always imply a high number of solved routes. For example, Chemformer has a higher solvability than MHNreact, yet produces a considerably lower number of solved routes per molecule (Chemformer: 8.04, MHNreact: 56.6). Moreover, a high number of explored routes is also not directly connected to a high solvability. For example, MHNreact explores the highest number of routes per molecule but performs only the third best in solvability since it solves comparably few explored routes.

Lastly, there are large disparities across the average search time per molecule. Chemformer is by far the slowest model (18737 sec), followed by MHNreact (8016 sec), both of which are considerably slower compared to LocalRetro (322 sec) and \gls{AZF} (129 sec). Even with these extensive search times, Chemformer and MHNreact do not reach the same level of single-step model calls as LocalRetro and \gls{AZF}.

Generally, LocalRetro outperforms other models in terms of solvability and number of solved routes while producing slightly fewer total explored routes than MHNreact and needing approximately 2.5x the time per molecule in comparison to the fastest \gls{AZF} baseline.

\newcommand{\ra}[1]{\renewcommand{\arraystretch}{#1}}
\begin{table*}[hbt]
\centering
\caption{\label{tab:ms}Comparison of multi-step algorithm and single-step retrosynthesis model combinations on USPTO-50k test set (5,002 molecules). Bold numbers indicate the best performance across all experiments.
}
\begin{tabular}{@{}crrcrrrr@{}}\toprule
& & \multicolumn{1}{c}{Overall} & \phantom{}& \multicolumn{4}{c}{Average per Molecule} \\
\cmidrule{3-3} \cmidrule{5-8}
\textbf{Algorithm} & \textbf{Model} & \begin{tabular}[c]{@{}l@{}} \textbf{Solvability (\%)} \end{tabular} && \begin{tabular}[c]{@{}l@{}}\textbf{Explored} \\ \textbf{Routes} \end{tabular} & \begin{tabular}[c]{@{}l@{}}\textbf{Solved} \\ \textbf{Routes}\end{tabular} & \begin{tabular}[c]{@{}l@{}}\textbf{Search} \\ \textbf{Time (s)}\end{tabular} & \begin{tabular}[c]{@{}l@{}}\textbf{Model} \\ \textbf{Calls} \end{tabular} \\ \midrule
\multirow{4}{*}{\begin{tabular}[c]{@{}l@{}}\gls{MCTS}\end{tabular}} & \gls{AZF} & 49.5 && 367 & 24.9 & 165 & 783 \\
& Chemformer & 44.3 && 4.40 & 3.33 & 2475 & 8.48\\
& LocalRetro & 71.5 && 86.7 & 27.4 & 1616 & 412\\
& MHNreact & 44.4 && 7.11 & 2.50 & 1842 & 29.5\\ \midrule
\multirow{4}{*}{\begin{tabular}[c]{@{}l@{}}Retro*\end{tabular}} & \gls{AZF} & 50.6 && 2574 & 48.6 & 130 & 195\\
& Chemformer & 53.4 && 39.7 & 2.06 & 1518 & 14.4\\  
& LocalRetro &  \textbf{80.6} && 7792 & 149 & 335 & 193\\
& MHNreact & 55.2 && 2818 & 17.6 & 1653 & 38.8\\ \midrule
\multirow{4}{*}{\begin{tabular}[c]{@{}l@{}}Retro*-extended\end{tabular}} & \gls{AZF} & 50.6 && 2567 & 48.5 &  \textbf{129} & 195\\
& Chemformer & 65.6 && 224 & 8.04 &  18738 & 134\\
& LocalRetro &  \textbf{80.6} && 7786 &  \textbf{151} & 322 & 193\\
& MHNreact & 60.9 &&  \textbf{8176} & 56.5 & 8016 & 180\\
\bottomrule
\end{tabular}
\end{table*}

\section{Discussion}

We show that a single-step retrosynthesis model can tremendously impact multi-step retrosynthesis planning, influencing the ability to solve products and successfully produce multiple solutions. Across all three experiments (\gls{MCTS}, Retro*, Retro*-extended), the alternative single-step models mostly outperform the baseline \gls{AZF} model. In \gls{MCTS}, one single-step model, LocalRetro, shows a considerably higher solvability than \gls{AZF}. In the case of Retro* and Retro*-extended, all models outperform \gls{AZF}, particularly when given an extended time to carry out sufficient model calls. The generally best performing model is LocalRetro, which has outstanding solvability and provides the most solved routes per molecule across all multi-step retrosynthesis planning experiments, continually outperforming all other methods. We show that the exchange of the single-step model alone can improve solvability by +30.0\%, reaching 80.6\%, and triple the number of solved routes per molecule to 151. As such, the single-step model should be well considered when developing multi-step retrosynthesis planning approaches.

Given our results, no clear pattern supports the usage of single-step top-k metrics as a potential proxy measure for solvability in multi-step retrosynthesis. For single-step models, the accuracy ranking varies from top-1 to top-50 (Fig. \ref{fig:accuracy} and Tab. \ref{tab:accuracy}). However, these rankings, and their intermediates, are never matched by their respective multi-step solvability rankings (Tab. \ref{tab:ms}). Hence, multi-step solvability does not solely depend on a singular single-step factor and should not be reduced to a singular top-k single-step metric.

Exclusively focusing on the top-1 accuracy of single-step models is especially problematic when transferring the model to a multi-step domain. Though Chemformer shows the highest top-1 accuracy, in multi-step experiments, it finds a comparatively low number of solved routes per molecule despite having the second highest solvability. A cause for this could be its single-step accuracy profile, going from the best performing to the worst performing model as top-k increases. This suggests that the model is proficient at predicting certain reactions but cannot find a diverse set of solutions. However, diverse solutions are beneficial for a tree-search setting, where up to 50 possible explorable route alternatives could be added per search iteration.

Importantly, similar top-k accuracy profiles do not result in the same multi-step results. For example, MHNreact and LocalRetro have a similar top-k accuracy profile in single-step retrosynthesis but differ greatly in multi-step retrosynthesis planning. Though they explore a similar number of routes, MHNreact solves considerably fewer routes. This difference in performance may be explained by comparing Retro* and Retro*-extended, where MHNreact improves only slightly in solvability despite having considerably more time. Though MHNreact explores and solves many more routes in general, the difference in solvability shows that it can only do this for molecules it had already solved in the shorter time frame.

Since multi-step solvability is not solely dependent on the top-k accuracy shown by the single-step models, other factors may contribute to their performance. For example, given that search algorithms generally have a limited run-time, single-step model inference times can greatly affect performance in multi-step retrosynthesis planning. To produce solved routes, single-step models must carry out as many single-step model calls as possible within an allocated time limit. If the inference time is too long, then the number of model calls will be limited, and as such, the number of explored routes will also be limited. This effect is evident in MHNreact and Chemformer, the models with the highest search times and lowest model calls across all experiments.

Single-step models typically evaluate in batches larger than one, however this does not currently transfer to multi-step retrosynthesis planning. For example, MHNreact has the fastest inference time when using its default batch size, but its inference time is considerably increased when reducing the batch size to one, the setting under which multi-step retrosynthesis planning is carried out.  As such, single-step models may not reach their full potential in search algorithms due to slow inference times. 

Noteworthy, most single-step models are developed for GPU use, and CPU usage can hinder their inference speed. However, it is necessary to conduct the multi-step experiments in parallel on CPUs due to the thousands of target molecules to be solved since massive GPU parallelization is currently not available to the general research community. Consequently, models designed and optimized for the single-step prediction task may not perform well in multi-step retrosynthesis planning. Therefore, single-step model developers should take the potential multi-step application into account and optimize their methods accordingly.

There are more general aspects to consider when discussing the divide between single-step and multi-step retrosynthesis. Though USPTO-50k is the most commonly used dataset for benchmarking single-step retrosynthesis prediction models, it represents only a limited area of the chemical space such that the models and our results may not apply to a more expansive chemical domain. Moreover, USPTO-50k comprises only single-step reactions, and the produced routes cannot be compared to reference routes to evaluate their validity. Recently, new benchmarks have emerged to address the lack of multi-step reference data \cite{genhedenPaRoutesFrameworkBenchmarking2022a}, so further work is required to quantify the produced routes. Ideally, one would assess these routes irrespective of a particular reference route since there are many potential valid routes for any target molecule. However, at present, this can only be addressed by a domain expert, an extremely time-intensive task. 

Additionally, this work focuses on using single-step models within two selected search algorithms. However, other search approaches can also considerably impact multi-step retrosynthesis planning methods \cite{hongRetrosyntheticPlanningExperienceGuided2021, kimSelfImprovedRetrosyntheticPlanning2021}. Thus finding the optimal combination of single-step and multi-step methods is yet to be explored and could have a substantial impact on synthesis planning in the future.

\section{Conclusion}

In this work, we bridge the gap between single-step retrosynthesis and multi-step retrosynthesis planning. By extending current state-of-the-art single-step models to the multi-step domain, we find no clear relationship between the single-step and multi-step benchmarking metrics, in particular single-step top-k accuracy and multi-step solvability. Additionally, we show the importance of developing single-step models for the multi-step domain, as single-step models can have an impressive impact on multi-step retrosynthesis planning performance. LocalRetro, the best performing single-step model, increases solvability by +30.0\% to 80.6\% and triples the number of solved routes compared to the most widely used model. Interestingly, LocalRetro outperforms other single-step models, even those with similar single-step accuracy profiles. Additionally, we analyze other potential factors involved in the translation between the two domains, most notably the inference time of the single-step model. Overall, we show there is no easy transfer of single-step retrosynthesis models to the multi-step retrosynthesis planning domain.

With this work, we provide an overview of how current state-of-the-art single-step models fare within contemporary search algorithms, however, we only evaluate a selected scope of single-step and multi-step combinations. In the future, more diverse chemical datasets need to be further explored to examine the applicability of these approaches beyond the USPTO-50k dataset.

To summarize, we show that single-step models should be developed and tested for the multi-step domain, and not as an isolated task, to successfully identify synthesis routes for molecules of interest. 

\section*{Acknowledgements}
This study was partially funded by the European Union’s Horizon 2020 research and innovation program under the Marie Skłodowska-Curie Innovative Training Network European Industrial Doctorate grant agreement No. 956832 “Advanced machine learning for Innovative Drug Discovery”. Parts of this work were performed using the ALICE compute resources provided by Leiden University. We thank Dr. Anthe Janssen (Leiden Institute of Chemistry) for providing chemical feedback.

\bibliographystyle{IEEEtran}
\bibliography{ai4science_ssm_benchmark}

\newpage
\section*{Appendix}
\setcounter{table}{0}
\renewcommand{\thetable}{A\arabic{table}}

\begin{table*}[thb]
\caption{\label{tab:ssm_hyperparam} Single-step model training hyperparameters. Values are obtained from the respective publication. Where available, models are trained using hyperparameters specifically for USPTO-50k. AZF uses default hyperparameters as described in \cite{thakkarDatasetsTheirInfluence2020}. LocalRetro uses default hyperparameters as described in \cite{chenDeepRetrosyntheticReaction2021a}.  
}
    \begin{subtable}[t]{0.45\textwidth}
        \centering
        \caption{Chemformer \cite{irwinChemformerPretrainedTransformer2022}}
        \label{tab:hyperparam_chemformer}
        \begin{tabular}{ll}
        \toprule
        \textbf{Hyperparameter} & \textbf{Value} \\
        \midrule
        dataset & uspto\_50 \\
        pre-trained model & chemformer \\
        task & backward prediction \\
        epochs & 100 \\
        learning rate & 0.001\\
        schedule & cycle\\
        batch size & 128\\
        acc batches & 4\\
        augment & all\\
        augment prob. & 0.5\\
        \bottomrule
       \end{tabular}
    \end{subtable}
    \hfill
    \begin{subtable}[t]{0.45\textwidth}
        \centering
        \caption{MHNreact \cite{seidlImprovingFewZeroShot2022}}
        \label{tab:hyperparam_mhn}
        \begin{tabular}{ll}
        \toprule
        \textbf{Hyperparameter} & \textbf{Value} \\
        \midrule
        batch size & 1024\\
        dropout & 0.4\\
        epochs & 40\\
        evaluation every n epochs & 10\\
        learning rate & 1e-4\\
        \\
        \textit{molecular encoder} \\
        fingerprint size & 3e4\\
        fingerprint type & MxFP\\
        \\
        \textit{template encoder 1} \\
        template fingerprint type & MxFP \\
        random template threshold & 1\\
        reactant pooling & lgamma \\
        \\
        \textit{template encoder 2} \\
        template fingerprint type & rdk \\
        \\
        \textit{Hopfield layer 1 and 2} \\
        beta & 0.03\\
        association-dimension \textit{d} & 1024\\
        hopf-num-layers & 2\\
        % layer 2 weight & 0.1 \\
        \bottomrule
       \end{tabular}
    \end{subtable}
    \hfill
\end{table*}

\begin{table*}[htb]
\caption{\label{tab:batch_sizes}Single-step model default inference batch size.
}
\centering
\begin{tabular}{llll}\toprule
\textbf{Hyperparameter} & \textbf{LocalRetro} & \textbf{Chemformer} & \textbf{MHNreact} \\ \midrule
Batch Size &  16 & 64 & 32 \\
\bottomrule
\end{tabular}
\end{table*}

\clearpage
\begin{table*}[htb]
\centering
\caption{\label{tab:ms_hyperparam}Search algorithm settings for multi-step experiments. Italics indicates difference across experiments.
}
        \begin{tabular}{llll}
        \toprule
        \textbf{Setting} & \textbf{\gls{MCTS}} & \textbf{Retro*} & \textbf{Retro*-extended} \\
        \midrule
        search algorithm & \textit{mcts} & \textit{retrostar} & \textit{retrostar}\\
        c & 1.4 & \multicolumn{1}{c}{-} & \multicolumn{1}{c}{-} \\
        cumulative cutoff & 0.995 & 0.995 & 0.995\\
        cutoff number & 50 & 50 & 50\\
        max transformations & 6 & 6 & 6\\
        default prior & 0.5 & 0.5 & 0.5\\
        use prior & True & True & True\\
        return first & False & False & False\\
        iteration limit & 200 & 200 & 200\\
        time limit & \textit{1800} & \textit{1800} & \textit{28800}\\
        exclude target from stock & True & True & True\\
        prune cycle in search & True & True & True\\
        additive expansion & False & False & False\\
        stock & zinc & zinc & zinc\\
        \bottomrule
       \end{tabular}
       
\end{table*}

\begin{table*}[htb]
\caption{\label{tab:accuracy}Percentage of compounds for which single-step retrosynthesis models found the ground-truth reactants within the top-k (Accuracy) on  USPTO-50k averaged across 10-fold cross-validation.
}
\centering
\begin{tabular}{llllll}\toprule
\textbf{Model} & \textbf{Top-1} & \textbf{Top-3} & \textbf{Top-5} & \textbf{Top-10} & \textbf{Top-50} \\ \midrule
\gls{AZF} & 43.3 ± 1.0  & 60.1 ± 1.0 & 67.7 ± 0.9 & 75.5 ± 0.7 & 78.1 ± 0.7 \\
LocalRetro & 52.5 ± 0.7 & \textbf{76.6 ± 0.6} & \textbf{84.5 ± 0.6} & \textbf{91.1 ± 0.5} & \textbf{96.6 ± 0.3} \\
Chemformer (beam size 50) & \textbf{54.7 ± 1.1} & 65.9 ± 1.0  & 67.9 ± 0.8 & 69.1 ± 0.8 & 73.3 ± 0.3 \\
Chemformer (beam size 10) & \textbf{54.7 ± 1.1} & 66.9 ± 0.9 & 69.4 ± 0.8 & 71.2 ± 0.7 & \multicolumn{1}{c}{-} \\
MHNreact & 49.8 ± 0.8  & 72.8 ± 1.0 & 80.2 ± 0.8 & 86.7 ± 0.6 & 93.3 ± 0.4 \\
\bottomrule
\end{tabular}
\end{table*}

\end{document}